\RequirePackage{lineno}

\documentclass[aps,prl,twocolumn,superscriptaddress,
showpacs,
preprintnumbers,
amsmath,amssymb,floatfix]{revtex4-2}
\usepackage{subcaption}

\usepackage{graphicx} 
\usepackage{dcolumn}  
\usepackage{amsmath}
\usepackage{orcidlink}

\graphicspath{{ps}}

\newcommand{\mevcsq}{\ensuremath{\textrm{MeV}/\textit{c}^2}}
\newcommand{\kevcsq}{\ensuremath{\textrm{keV}/\textit{c}^2}}

\let\oldequation\equation
\let\oldendequation\endequation

\renewenvironment{equation}
  {\linenomathNonumbers\oldequation}
  {\oldendequation\endlinenomath}

\begin{document}



\title{ \quad\\[1.0cm] Measurement of the mass and width
of the $\Lambda_c(2625)^+$ charmed baryon and the branching ratios of
$\Lambda_c(2625)^+ \to \Sigma_c^{0}\pi^{+}$ and 
$\Lambda_c(2625)^+ \to \Sigma_c^{++}\pi^{-}$}

\noaffiliation
  \author{D.~Wang\,\orcidlink{0000-0003-1485-2143}} 
  \author{J.~Yelton\,\orcidlink{0000-0001-8840-3346}} 
  \author{I.~Adachi\,\orcidlink{0000-0003-2287-0173}} 
  \author{J.~K.~Ahn\,\orcidlink{0000-0002-5795-2243}} 
  \author{H.~Aihara\,\orcidlink{0000-0002-1907-5964}} 
  \author{D.~M.~Asner\,\orcidlink{0000-0002-1586-5790}} 
  \author{H.~Atmacan\,\orcidlink{0000-0003-2435-501X}} 
  \author{R.~Ayad\,\orcidlink{0000-0003-3466-9290}} 
  \author{V.~Babu\,\orcidlink{0000-0003-0419-6912}} 
  \author{Sw.~Banerjee\,\orcidlink{0000-0001-8852-2409}} 
% ...
\collaboration{The Belle Collaboration}



\begin{abstract}
Using the entire data sample of $980\,\textrm{fb}^{-1}$
collected at or near the $\Upsilon(4S)$ resonance
with the Belle detector operating at the KEKB
asymmetric-energy $e^{+}e^{-}$ collider 
, we report the measurement 
of the mass, width, and the branching 
ratios of the $\Lambda_c(2625)^+$ charmed baryon.
The mass difference between $\Lambda_c(2625)^+$ and 
$\Lambda_c^+$ is measured to be
$M(\Lambda_c(2625)^{+}) - M(\Lambda_c^{+}) = 
341.518 \pm 0.006 \pm 0.049\ \mevcsq$. 
The upper limit on the width is measured to be
$\Gamma(\Lambda_c(2625)^+) < 0.52\,\mevcsq$ 
at 90\% confidence level.
Based on a full Dalitz plot fit, branching ratios
with respect to the mode $\Lambda_c(2625)^+ \to \Lambda_c^+ \pi^+ \pi^-$
are measured to be
$\frac{\mathcal{B}(\Lambda_c(2625)^+ \to \Sigma_c^{0} \pi^{+})}
{\mathcal{B}(\Lambda_c(2625)^+ \to \Lambda_c^+ \pi^{+} \pi^{-})} = 
(5.19 \pm 0.23 \pm 0.40) \%$
and 
$\frac{\mathcal{B}(\Lambda_c(2625)^+ \to \Sigma_c^{++} \pi^{-})}
{\mathcal{B}(\Lambda_c(2625)^+ \to \Lambda_c^+ \pi^{+} \pi^{-})} 
= (5.13 \pm 0.26 \pm 0.32) \%$,
where the first and second uncertainties are statistical and
systematic, respectively.
These measurements can be used to further constrain the
parameters of the underlying theoretical models.

\end{abstract}

\maketitle

\tighten

{\renewcommand{\thefootnote}{\fnsymbol{footnote}}}
\setcounter{footnote}{0}

\section{Introduction}

The $\Lambda_c^+$ charmed baryons consist of a heavy charm quark
and two light ($ud$) quarks with the ground state 
having quantum numbers $J^P = \frac{1}{2}^+$.
The $\Lambda_c(2595)^+$ and $\Lambda_c(2625)^+$ are the two 
lowest-lying excited states observed, and are generally believed to have
$J^P = \frac{1}{2}^-$ and $J^P = \frac{3}{2}^-$, respectively. 
The $\Lambda_c(2595)^+$ predominantly decays
to the $J^P=\frac{1}{2}^+$ $\Sigma_c(2455)^{++/0}$ states
via an $s$-wave decay.
The analogous decay for the $\Lambda_c(2625)^+$ to the 
$J^P = \frac{3}{2}^+$  $\Sigma_c(2520)^{++/0}$ states 
is kinematically suppressed as it can only happen 
through the low-mass tail of the $\Sigma_c(2520)^{++/0}$.
The $d$-wave decay to the $J^P=\frac{1}{2}^+$  
$\Sigma_c(2455)^{++/0}$  states
is allowed, but its contribution is known to be small. 
Thus, the $\Lambda_c(2625)^+$ decay is thought to proceed
primarily via the
direct three-body, $p$-wave decay 
$\Lambda_c(2625)^+ \to \Lambda_c^+ \pi^+ \pi^-$.

The $\Lambda_c(2625)^+$ was first observed in 1993~\cite{ARGUS:1993}.
The CDF collaboration reported the most recent measurements of 
$\Lambda_c(2625)^+$ properties in 2011 using a data sample of 6.2k events~\cite{CDF:2011}.
Their measurement for the $\Lambda_c(2625)^+$ mass with respect to 
the $\Lambda_c^+$ mass is much more precise compared with previous
measurements, and an upper limit on the $\Lambda_c(2625)^+$ width
was reported. The limited decay phase space of 
$\Lambda_c(2625)^+ \to \Lambda_c^+ \pi^+ \pi^-$ makes it difficult 
to extract the $\Sigma_c(2455)^{++/0}$ yields by fitting the 
$\Lambda_c^+\pi^{\pm}$ invariant mass due to the presence of 
reflection peaks formed by the combination of the $\Lambda_c^+$ and 
the other final state pion. 
The large data sample collected by Belle, together with the use of an 
amplitude model~\cite{Arifi:2018prd} to describe the decay, allows 
us to use a full Dalitz
fit that naturally includes the reflections.

The mass of the $\Lambda_c(2625)^+$, relative to the $\Lambda_c^+$ mass,
is already relatively well known, but the large  Belle data sample 
allows for a more precise measurement.
No intrinsic width of the $\Lambda_c(2625)^+$ has yet been measured, 
and the current upper limit
$\Gamma < 0.97\ \mevcsq$ at 90\% confidence level
by the Particle Data Group (PDG)~\cite{PDG:2020} is based on the CDF measurement.

Theoretical predictions for the width vary 
for this narrow state~\cite{Arifi:2018prd, Arifi:2021prd, Kawakami:2018prd, Guo:2019prd}.
An improved limit on the width of the $\Lambda_c(2625)^+$ 
will help to constrain these predictions, 
and provide insights into other charmed baryons since 
their widths are related through common coupling constants~\cite{Cheng:2015}.

%
%
%

\section{Detector and dataset}

The measurement presented here is based on the entire dataset collected by the 
Belle detector~\cite{Belle, Belle:achievments} operating at the KEKB
asymmetric-energy $e^+ e^-$ collider~\cite{KEKB, KEKB:achievements}. 
The total integrated luminosity of the dataset is $980\ \textrm{fb}^{-1}$,
which is mostly collected at or near the $\Upsilon(4S)$ resonance. 

The Belle detector is a large-solid-angle magnetic
spectrometer that consists of a silicon vertex detector (SVD),
a 50-layer central drift chamber (CDC), an array of
aerogel threshold Cherenkov counters (ACC),  
a barrel-like arrangement of time-of-flight
scintillation counters (TOF), and an electromagnetic calorimeter
composed of CsI(Tl) crystals (ECL) located inside 
a super-conducting solenoid coil that provides a 1.5~T
magnetic field.  An iron flux-return located outside of
the coil is instrumented to detect $K_L^0$ mesons and to identify
muons (KLM).  The detector
is described in detail elsewhere~\cite{Belle}.
Two inner detector configurations were used. 
The first consisted of a 2.0 cm radius beampipe
and a 3-layer silicon vertex detector,
while the second used a 1.5 cm radius beampipe, a 4-layer
silicon detector and a small-cell inner drift chamber.  

Monte Carlo (MC) events are generated using EVTGEN~\cite{Evtgen}
to optimize selection criteria and to be used in the Dalitz plot fit.
The $\Lambda_c(2625)^+ \to \Lambda_c^+ \pi^+ \pi^-$ and 
$\Lambda_c^+ \to p K^- \pi^+$ samples are 
generated using a phase space model~\cite{charge_conjugates}.
A $D^{*+} \to D^0 \pi^+$, with $D^0 \to K^- \pi^+$, sample is also generated
to compare the mass-resolution function in the MC sample and the experimental data, and thus to estimate
the systematic uncertainties on the measurements.
The detector response is simulated with GEANT3~\cite{Geant3}
and the event reconstruction is performed using 
data converted with the Belle-to-Belle-II (B2BII) 
software package~\cite{B2BII}
and then analyzed using Belle II software~\cite{BASF2, basf2-zenodo}.


\section{Analysis}

The candidate $\Lambda_c(2625)^+$ baryons are reconstructed 
from the decay chain
$\Lambda_c(2625)^+ \to \Lambda_c^+ \pi^+ \pi^-$, $\Lambda_c^+ \to p K^- \pi^+$.
The final-state charged particles, $\pi^\pm$, $K^-$ and $p$, are selected
based on the likelihood information from the tracking (SVD, CDC) and
particle identification (CDC, ACC, TOF) systems into a combined likelihood,
$\mathcal{L}(h_1 {:} h_2) = \mathcal{L}(h_1) / (\mathcal{L}(h_1) + \mathcal{L}(h_2))$, 
where $h_1$ and $h_2$ are $p$, $K$ or $\pi$~\cite{BellePID}.
We require the proton candidates to have 
$\mathcal{L}(p{:}K) > 0.6$ and $\mathcal{L}(p{:}\pi) > 0.6$, 
kaon candidates to have $\mathcal{L}(K{:}p) > 0.6$ and $\mathcal{L}(K{:}\pi) > 0.6$, 
and pion candidates to have $\mathcal{L}(\pi{:}K) > 0.6$ and $\mathcal{L}(\pi{:}p) > 0.6$.
Electrons are suppressed by requiring 
$\mathcal{L}(e^-)/(\mathcal{L}(e^-) + \mathcal{L}(\mathrm{hadrons})) < 0.1$
for all candidates; the likelihoods $\mathcal{L}(e^-)$ and 
$\mathcal{L}(\mathrm{hadrons})$ include information from the ECL 
in addition to the tracking and particle identification systems~\cite{BellePID}.
The particle identification efficiency is approximately 
87\% for protons, 85\% for kaons and
96\% for pions.
Charged tracks are also required to have a 
point of closest approach
with respect to the interaction point 
less than 3 cm in the $e^{+}$ beam direction
and less than 1 cm in the plane perpendicular to it.

A vertex fit is applied to the daughter particles of 
the $\Lambda_c^+$ candidates and 
the resultant $\chi^2$ probability of the fit 
is required to be greater than 0.001. 
Candidates within $\pm 7.0\ \mevcsq\ (\approx 1.6\sigma)$ 
are selected and mass-constrained 
to the $\Lambda_c^+$ PDG mass of $2286.46\,\mevcsq$ ~\cite{PDG:2020}. 
Two pions of opposite charge are then combined 
with the constrained $\Lambda_c^+$ candidate to form a $\Lambda_c(2625)^+$
candidate. The $\Lambda_c(2625)^+$ daughters are then kinematically fitted to come from 
a common vertex, with a constraint that the vertex has to be within the beamspot since the $\Lambda_c(2625)^+$ is short-lived. 
The $\chi^2$ probability of this fit is required to be 
greater than 0.001 to ensure the quality of the fit. 
As excited charmed baryons including the $\Lambda_c(2625)^+$ typically have
a hard momentum distribution, we only keep $\Lambda_c(2625)^+$ candidates
with $x_p > 0.7$, where 
$x_p = p^{*} / \sqrt{E_{\mathrm{beam}}^2/\mathit{c}^2 - M^2\mathit{c}^2}$ and
$p^*$ is momentum of the $\Lambda_c(2625)^+$ 
in the $e^+e^-$ center of mass frame.
As the mass of the $\Lambda_c^+$ is constrained to its PDG
value, the reconstructed mass $M(\Lambda_c^+\pi^+\pi^-)$
has the resolution of the
mass difference $M(\Lambda_c(2625)^+) - M(\Lambda_c^+)$.

Correctly calibrating the momentum scale for low momentum pions
is critical for this analysis. 
We calibrate the momentum
scale using copious $K^0_S\to\pi^+\pi^-$ events in the experimental data. 
Low-momentum tracks are iteratively calibrated as a 
function of the polar angle
and momentum of each track in the laboratory frame 
by comparing the reconstructed and world-average mass of the $K^0_S$
meson as a function of the $K^0_S$ momentum. 
This correction has been used in a previous $\Sigma_c^{++/0}$
study using Belle data~\cite{SLee:2014PRD}. 
Since the mass-resolution function is 
crucial for the precise measurement of 
the $\Lambda_c(2625)^+$ mass and width,
the MC tracks are smeared using
the analysis software during reconstruction, 
as otherwise the MC mass resolution is known
to be better than that of the experimental data. 
This track smearing affects
the width of the mass-resolution function but not
its central value.
The mass-resolution function of the $\Lambda_c(2625)^+$ mass is parameterized
as a sum of two Gaussian functions with parameters fixed according to
a signal MC sample with both corrections as detailed above.

The consistency between the MC sample and the experimental data 
is checked by comparing the mass resolution of $D^{*+}$ events, 
which have similar
kinematics to the events under study.
The low-momentum track correction ensures that the
measured $D^{*+}$ mass in data and MC are independent of the
momentum of the soft pion~\cite{SLee:2014PRD}.
The resolution of the $D^{*+}$ mass relative to the $D^{0}$ mass
in the experimental data is found by fitting the $M(D^0\pi^+) - M(D^0)$
mass distribution in the experimental data with a Breit-Wigner distribution
convolved with a double-Gaussian mass-resolution function, 
where the width
of the Breit-Wigner is fixed to the PDG value of $83.4\ \kevcsq$~\cite{PDG:2020}.
In this study, without track smearing, the mass resolution
in the experimental data is measured to be 114\% of the value obtained from the
MC sample.
However, with track smearing, 
the mass resolution
in the experimental data is measured to be
86\% of the value obtained from the MC sample. 
In all other narrow signals studied, for instance the $\Lambda_c^+$, the track smearing
ensures that the MC and data agree reasonably. 
The track smearing has negligible effect on the mass
measurement.
The results of these consistency checks are used in the estimation of the systematic
uncertainties described below. 

\begin{figure}[htb]
\includegraphics[width=\linewidth]{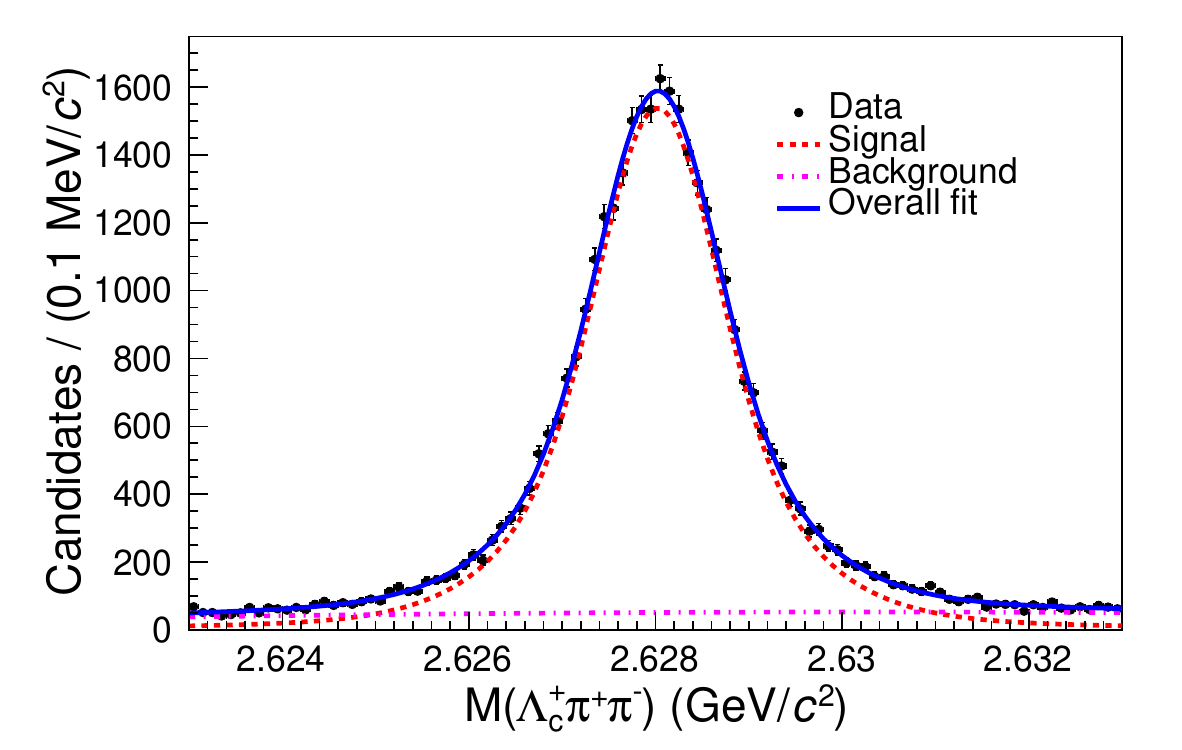}
\caption{Distribution of the invariant mass $M(\Lambda_c^+\pi^+\pi^-)$
where the $\Lambda_c^+$ mass is fixed to the PDG value.
The solid line shows the overall distribution and the dashed lines
show the individual signal and background components.
}
\label{mass_fit1}
\end{figure}

The reconstructed $M(\Lambda_c^+\pi^+\pi^-)$ mass distribution 
in the experimental data is fitted using RooFit~\cite{Roofit}.
Figure~\ref{mass_fit1} shows the $M(\Lambda_c^+\pi^+\pi^-)$ mass distribution
in the experimental data overlaid with the fit result.
The signal function is a Breit-Wigner distribution 
convolved with a double-Gaussian mass-resolution function, and
the background function is a second-order Chebychev polynomial.
The resolution function for the invariant mass distribution
is obtained from the MC sample, without track smearing, and
scaled by 114\% in accordance with the $D^{*+}$ study.
The solid line shows the overall fit and the dashed lines 
show the individual signal and background components of the fit.
The fitted mass is $2628.025 \pm 0.006 \,\mevcsq$, independent of
which version of the mass-resolution function we use.
The uncertainty is statistical only.
If we use the track-smearing correction without any rescaling,
the fitted width is found to be zero, so we have no definitive 
evidence of a non-zero width and will present only an upper limit
for the measurement of the intrinsic width of the $\Lambda_c^+(2625)$.
If we scale by 114\% the mass-resolution function without track smearing,
the fitted width is $0.490 \pm 0.025 \,\mevcsq$. 
If we scale by 86\% the mass-resolution function with track smearing, 
the fitted width is $0.293 \pm 0.026 \,\mevcsq$.
These finite values for the fitted width after scaling the mass resolution
are only used to find the limit on the intrinsic width including
systematic uncertainties.

The fitted mass of $\Lambda_c(2625)^+$ in the signal
MC sample is slightly different from the generated value.
Applying a bias correction, determined by the mass shift observed
in the signal MC sample, 
the mass of the $\Lambda_c(2625)^+$ is measured to be 
$2627.978 \pm 0.006 \,\mevcsq$,
where the uncertainty is statistical.

Two upper limits on the width are calculated 
based on the two mass-resolution functions methods described above 
and the larger upper limit 
is reported as the final answer. 
Using the mass-resolution function determined from MC 
scaled by 114\% without
track smearing, the upper limit
is determined to be
\begin{equation}
\Gamma(\Lambda_c(2625)^+) < 0.52 \,\mevcsq
\end{equation}
at 90\% confidence level by integrating the likelihood function 
to find the value for which the integral contains
90\% of the total area.
Using the mass-resolution function scaled by 
86\% with track smearing 
would yield a tighter upper limit. 
Therefore, we conservatively report the former as 
the upper limit on the width of $\Lambda_c(2625)^+$.

\section{Dalitz plot fit}
A Dalitz plot fit is made in order to determine the branching 
ratios of $\Lambda_c(2625)^+$ with respect to the mode
$\Lambda_c(2625)^+ \to \Lambda_c^+ \pi^+ \pi^-$.
For the Dalitz plot fit, only $\Lambda_c(2625)^+$ candidates 
within $\pm 2\,\mevcsq$ of the $\Lambda_c(2625)^+$ PDG mass are accepted~\cite{PDG:2020}.
The invariant mass of the $\Lambda_c(2625)^+$ candidates
is then constrained to the $\Lambda_c(2625)^+$  PDG mass of $2628.11\,\mevcsq$,
and the four-vectors of 
the daughter particles are updated accordingly.
A fit is made to the Dalitz plot using an amplitude model
as presented by Arifi {\em et al.}~\cite{Arifi:2018prd} 
using the AmpTools software package~\cite{AMPTOOLS}.
The $\Lambda_c(2625)^+$ signal distribution is calculated
from the squared amplitude with spin sum of final states and 
spin average of the initial states
\begin{equation}
\sum |\mathcal{T}_1 + \mathcal{T}_2 + \mathcal{T}_3 + 
\mathcal{T}_4 + \mathcal{T}_5|^2    
\end{equation}
where $\mathcal{T}_1$ through $\mathcal{T}_5$ are the decay 
amplitudes through the intermediate states
$\Sigma_c^0$, $\Sigma_c(2520)^0$, 
$\Sigma_c^{++}$, $\Sigma_c(2520)^{++}$, and the 
direct three-body decay, respectively.
Each amplitude is modeled as a Breit-Wigner function multiplied
by a form factor specific to each decay channel.
A constant amplitude is used to model the background 
$\Lambda_c^+\pi^+\pi^-$ combinations, which are not decay products
of $\Lambda_c(2625)^+$. The yield of each decay channel is calculated
using AmpTools by an integration of the individual component
over the Dalitz plot.
The contribution of the three-body decay in the signal model
is different from the background phase space decay in that
the former is not flat across the Dalitz plot.
During the fit, 
the masses and widths of these intermediate particles are constrained
to their respective PDG values to facilitate the convergence.
The small variations of the detector acceptance across the Dalitz plot
are taken into account by using the output of a phase space 
MC sample passed through the GEANT3 detector simulation 
as input to the AmpTools fitting package.

\begin{figure*}[htb]
\begin{subfigure}[b]{0.49\linewidth}
\includegraphics[width=\linewidth]{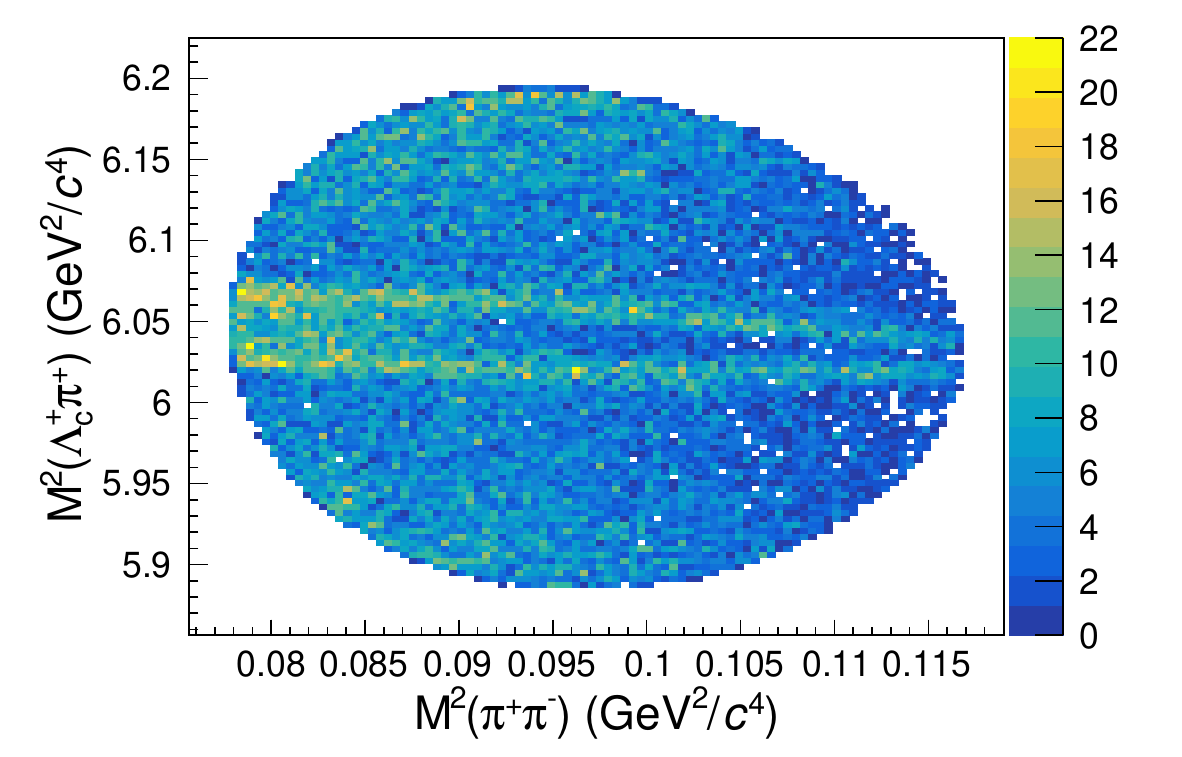}
\end{subfigure}
\begin{subfigure}[b]{0.49\linewidth}
\includegraphics[width=\linewidth]{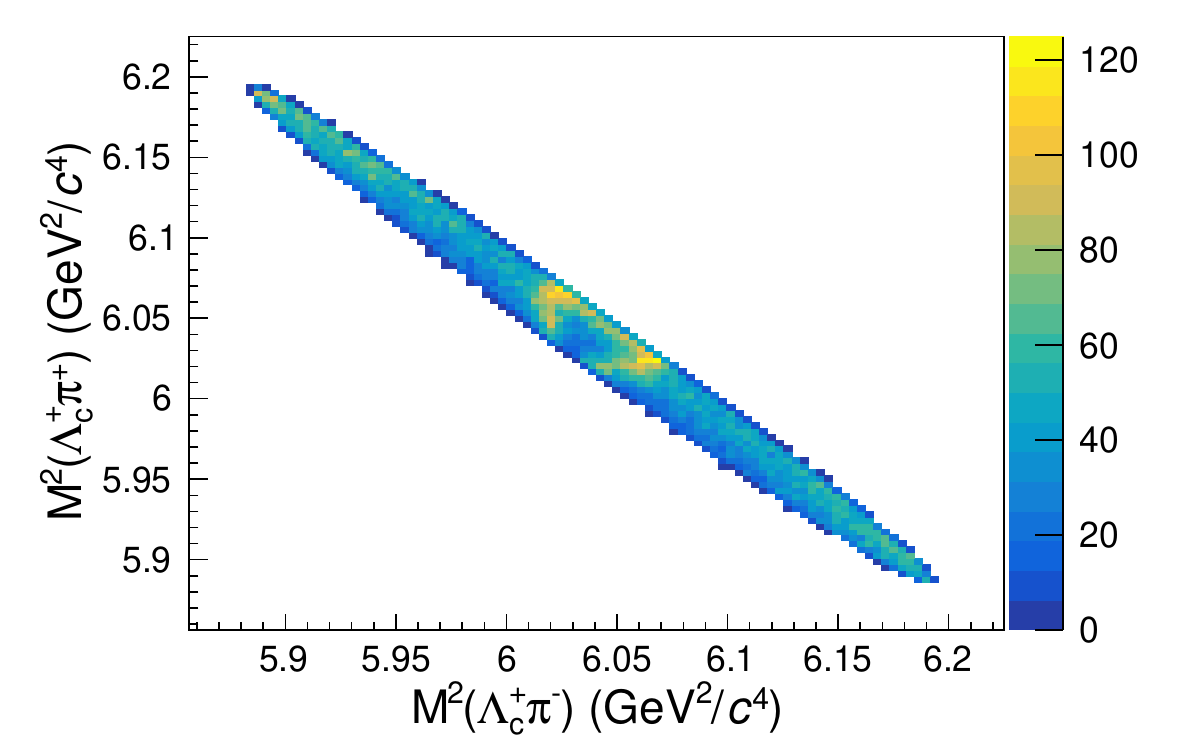}
\end{subfigure}
\caption{Dalitz plot for $\Lambda_c(2625)^+$ candidates in the signal region. Explanations of the patterns in the text.}
\label{data_dalitz}
\end{figure*}

Figure~\ref{data_dalitz} shows the Dalitz plots for
candidates in the signal region. 
On the left subplot, the contributions
from $\Sigma_c^{++}$ and the reflection 
from $\Sigma_c^0$ constitute
the two horizontal stripes. The upper and lower parts
of the Dalitz plot show slight excesses
due to $\Sigma_c(2520)^{++/0}$ decays. There is also 
a clear excess on the left side of the Dalitz plot compared to
the right in agreement with the three-body decay taking into account the spin, 
as predicted in the amplitude 
model~\cite{Arifi:2018prd}.
On the right subplot, the horizontal and vertical stripes
indicate the $\Sigma_c^{++}$ and $\Sigma_c^0$ decays
respectively. It is straightforward to see the origin of 
the reflection peaks on the $M(\Lambda_c^+\pi^+)$ mass projection
from this 2D Dalitz plot.

\begin{figure*}[htb]
\begin{subfigure}[b]{0.49\linewidth}
\includegraphics[width=\linewidth]{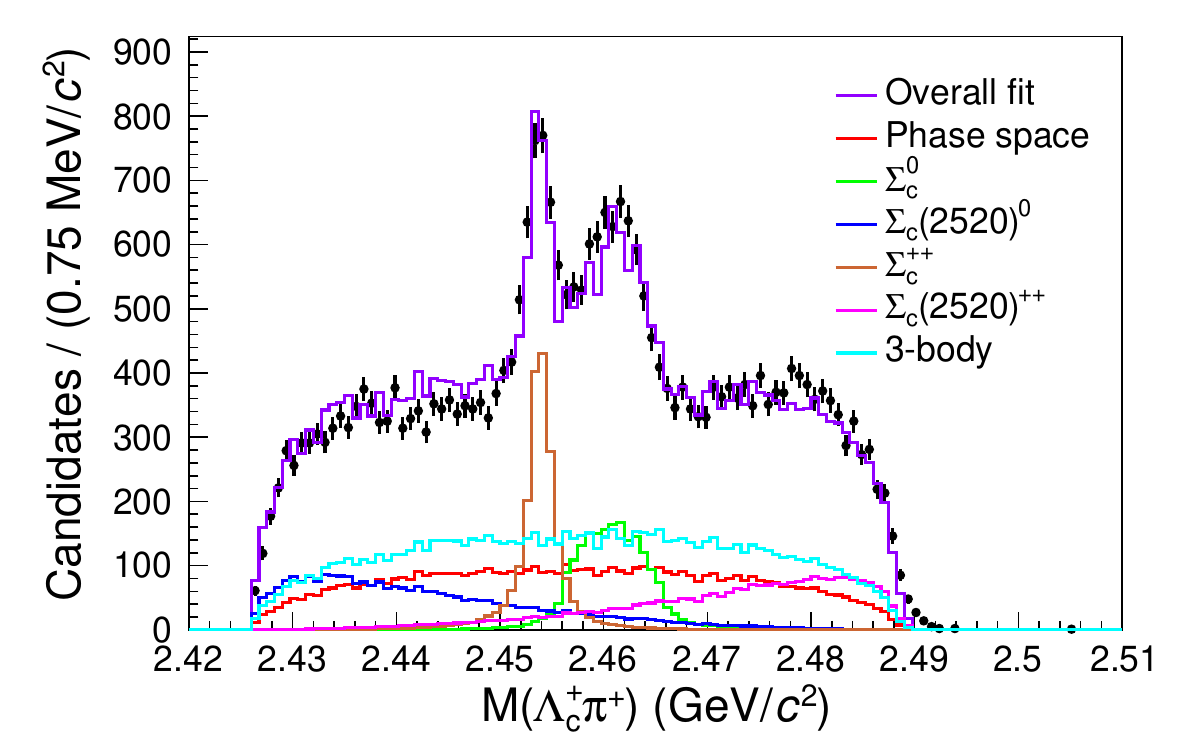}
\end{subfigure}
\begin{subfigure}[b]{0.49\linewidth}
\includegraphics[width=\linewidth]{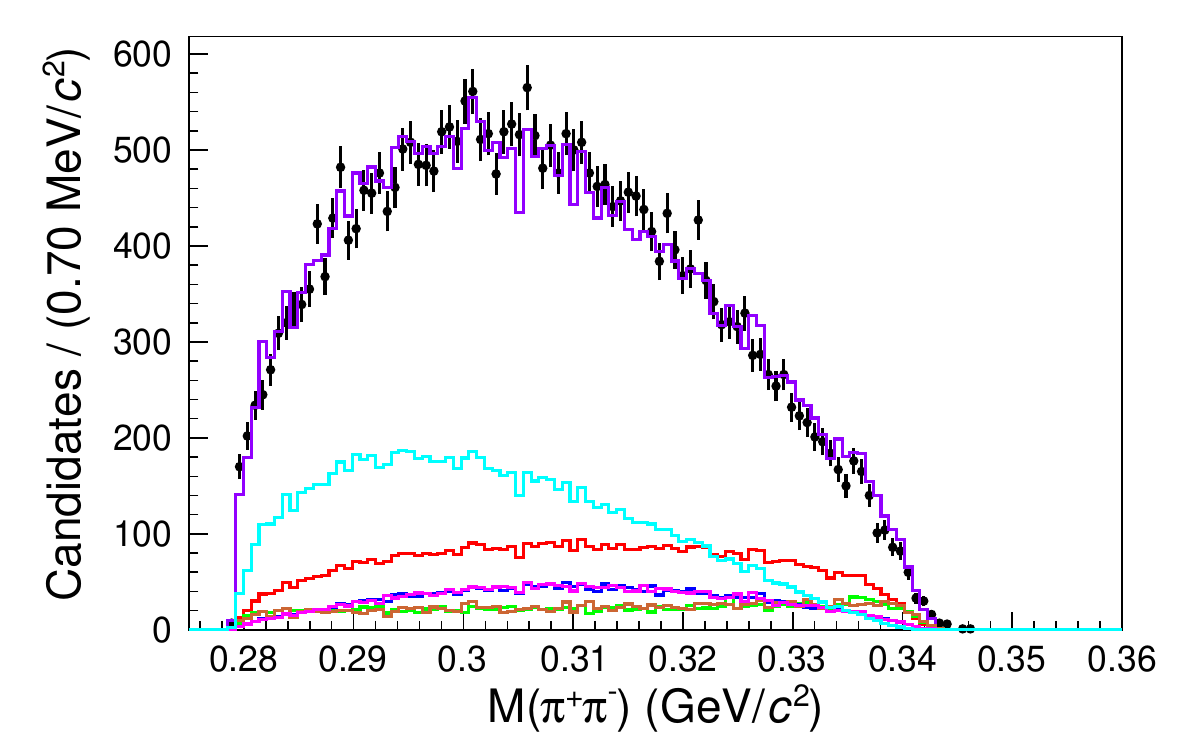}
\end{subfigure}
\caption{Dalitz plot fit result plotted as projections.
Solid lines show the overall fitted distribution and its
individual components as indicated in the legend.
More explanations in the text.}
\label{data_dalitz_projection}
\end{figure*}

Figure~\ref{data_dalitz_projection} shows the projections
of the fitted results with each component labeled on the plot.
The $\Sigma_c^{++}$ peak and the reflection peak 
from $\Sigma_c^0$ are evident on the $M(\Lambda_c^+\pi^+)$ mass
projection.
The shoulders on the left and right side of the mass region
are mostly formed by the decays from 
the off-shell $\Sigma_c(2520)^{++/0}$.
The three-body $p$-wave decay in the signal model shows up
in the $M(\pi^+\pi^-)$ mass projection as 
an asymmetric distribution, in contrast to 
the symmetric distribution from the background
phase space decay.
The $\Lambda_c(2625)^+$ yield in the signal region is
$N_{\mathrm{sig}}(\Lambda_c(2625)^+)) = 30319 \pm 371$. 
The  $\Sigma_c^{0}$ yield is 
$N_{\mathrm{sig}}(\Sigma_c^{0}) = 1964 \pm 66$
and the $\Sigma_c^{++}$ yield is 
$N_{\mathrm{sig}}(\Sigma_c^{++}) = 2022 \pm 76$.
The uncertainties are statistical only.

\begin{figure}[htb]
\includegraphics[width=\linewidth]{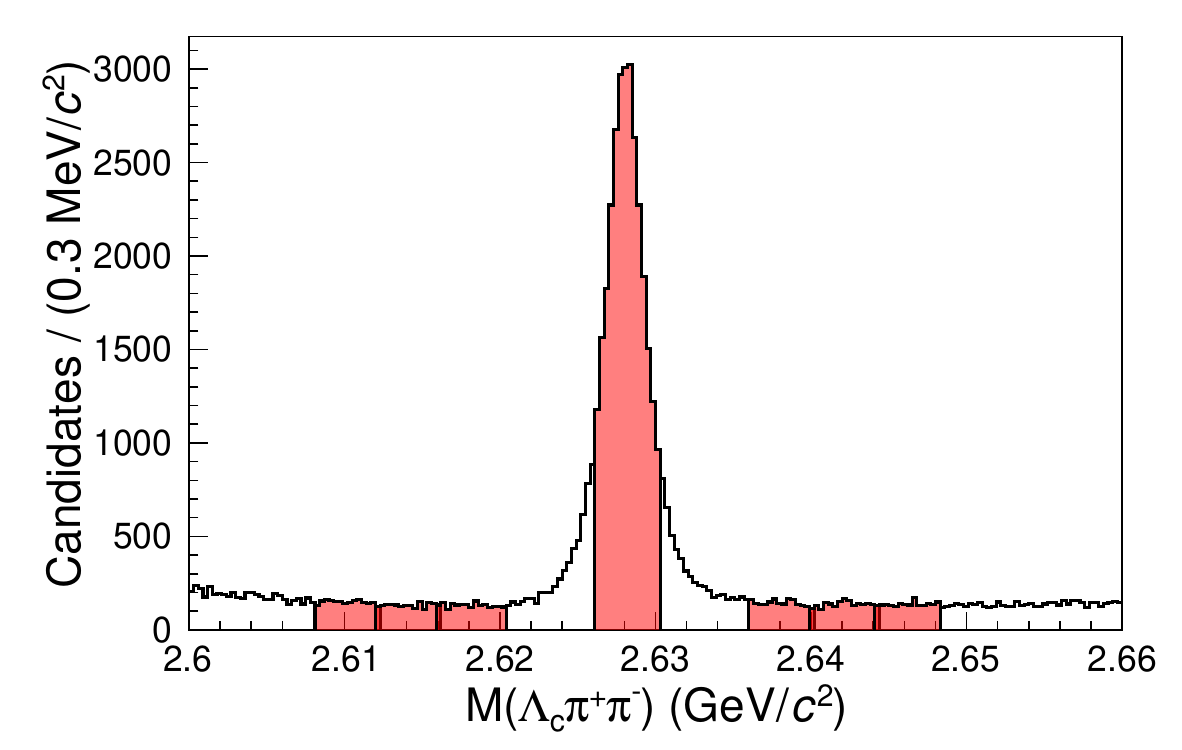}
\caption{Signal region and the six sideband regions on either
side of the signal region used for sideband subtraction.}
\label{fig:region_def}
\end{figure}

\begin{figure*}[htb]
\includegraphics[width=0.48\linewidth]{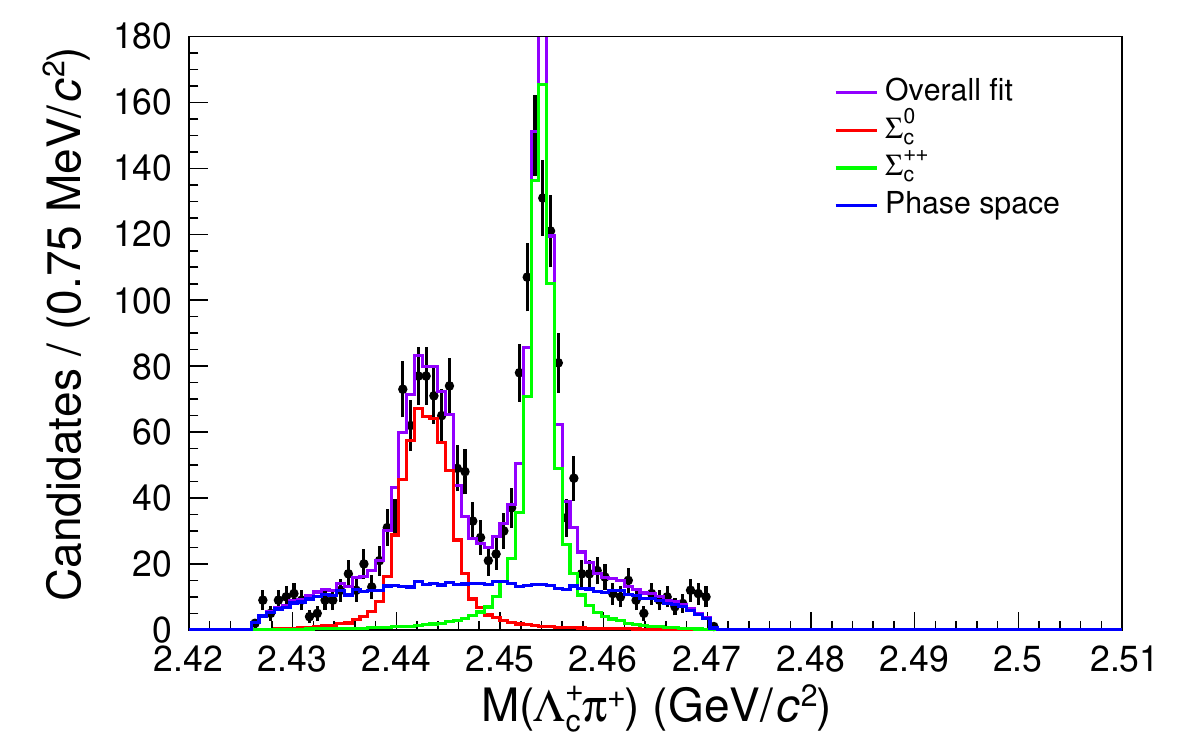}
\includegraphics[width=0.48\linewidth]{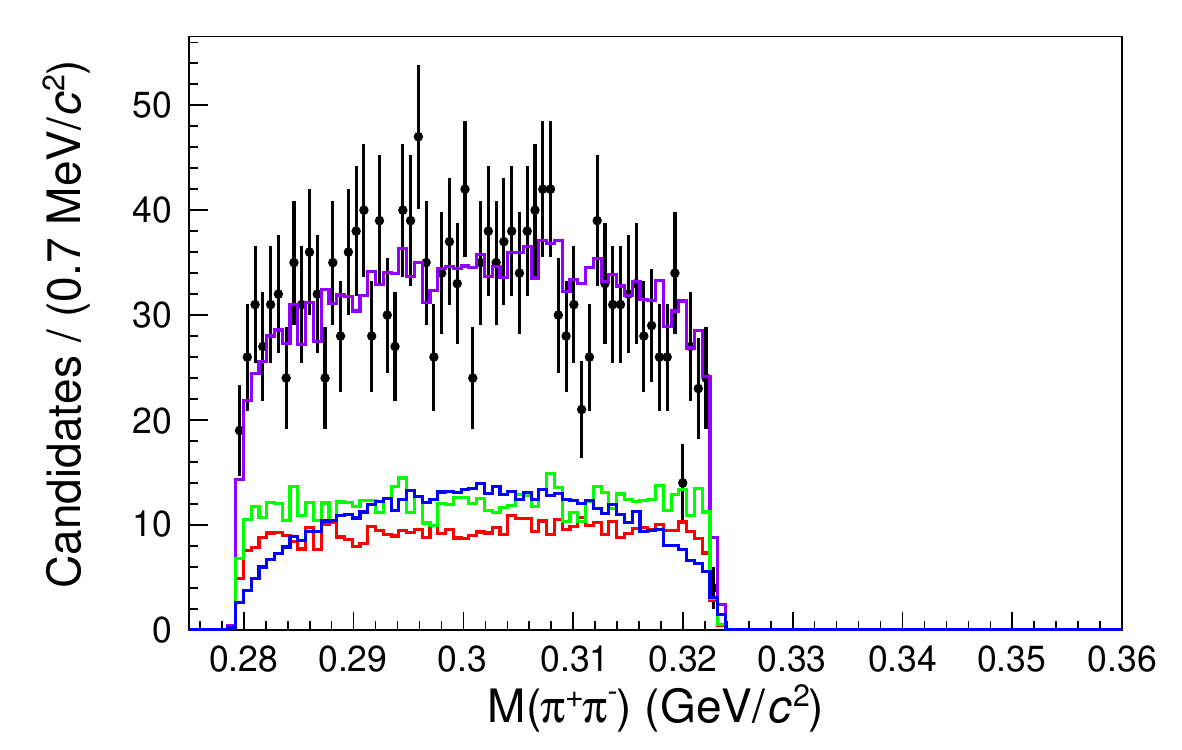} 
\includegraphics[width=0.48\linewidth]{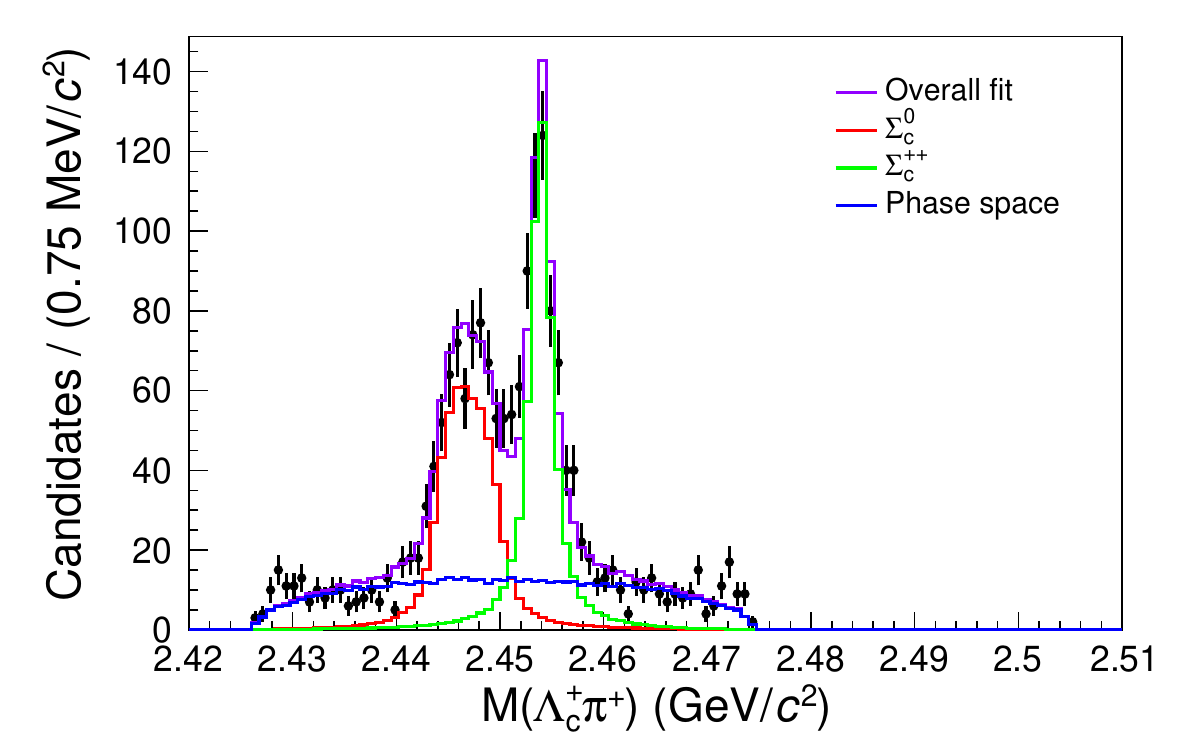}
\includegraphics[width=0.48\linewidth]{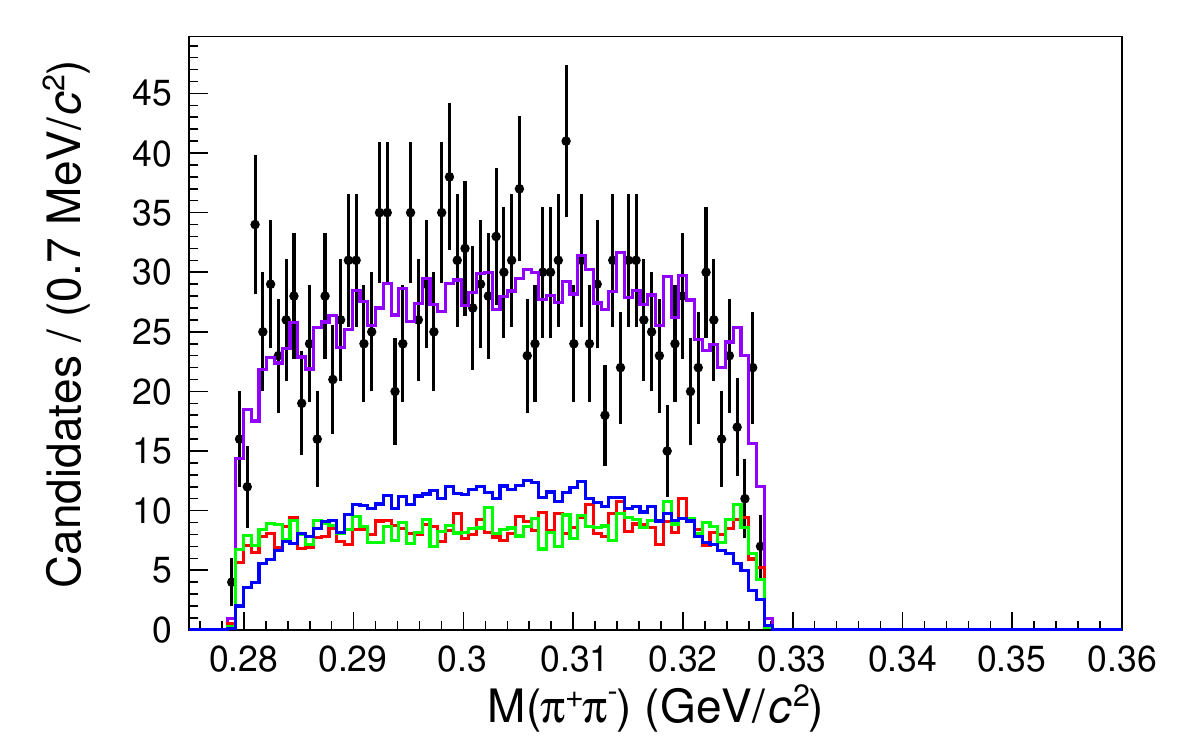} 
\includegraphics[width=0.48\linewidth]{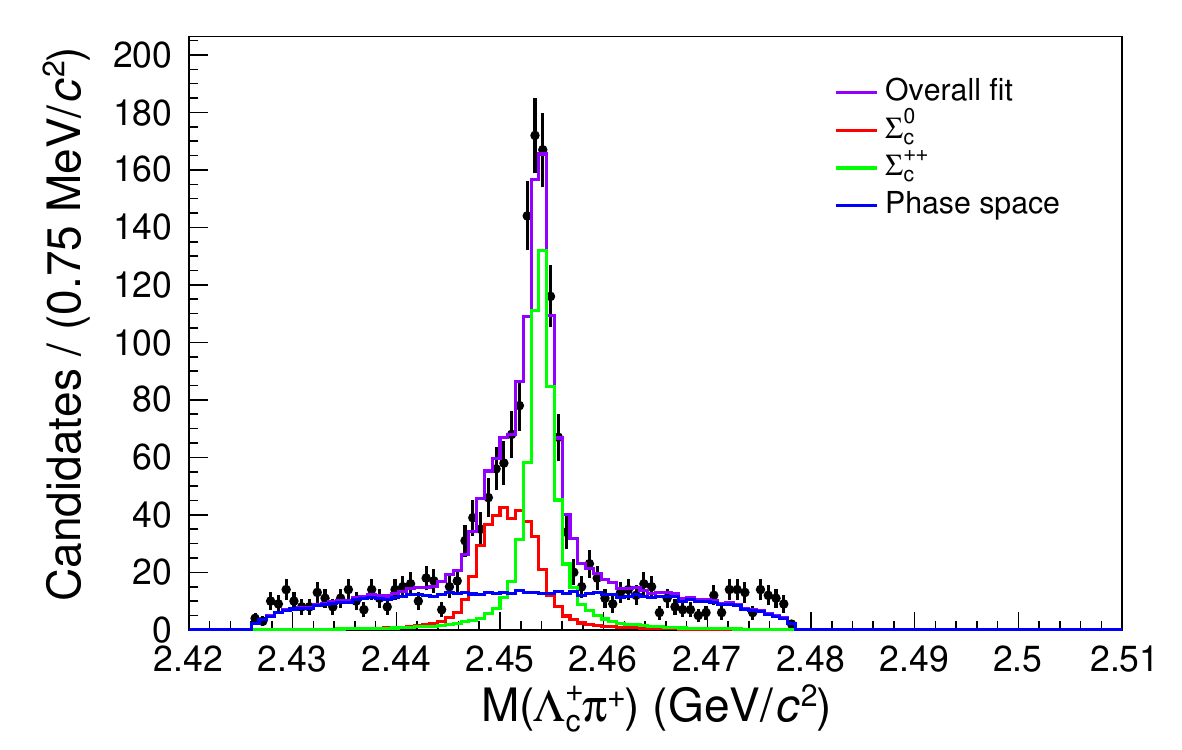}
\includegraphics[width=0.48\linewidth]{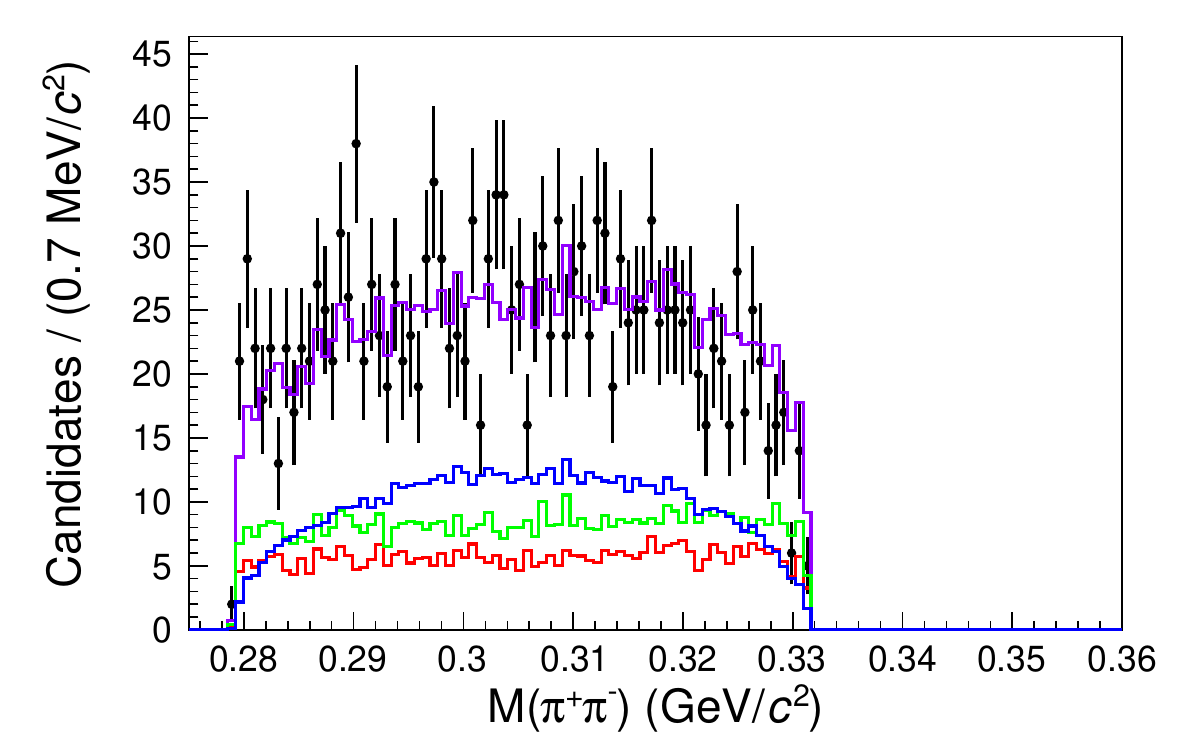}
\caption{Projections of the Dalitz plot fits of the 3 sidebands on the left side of the signal region. The overall fitted distribution and the individual fitted components are shown alongside the experimental data.}
\label{fig:sideband_dalitz_left}
\end{figure*}

\begin{figure*}[htb]
\includegraphics[width=0.48\linewidth]{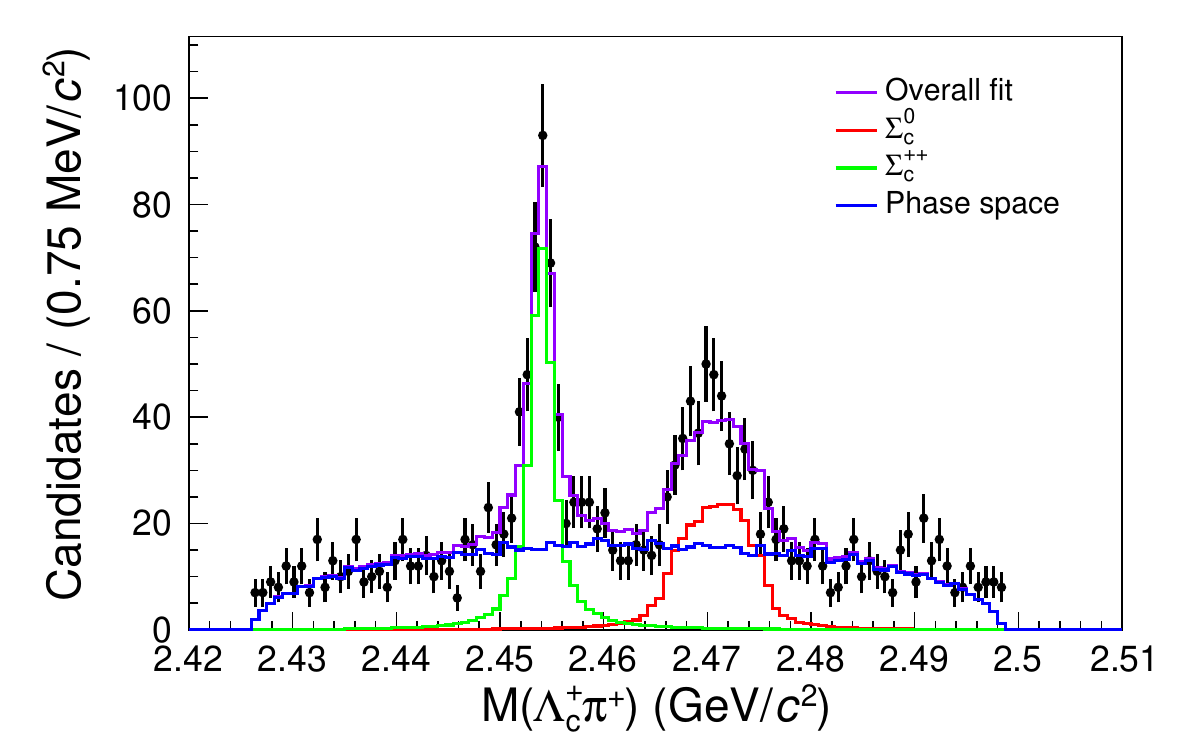}
\includegraphics[width=0.48\linewidth]{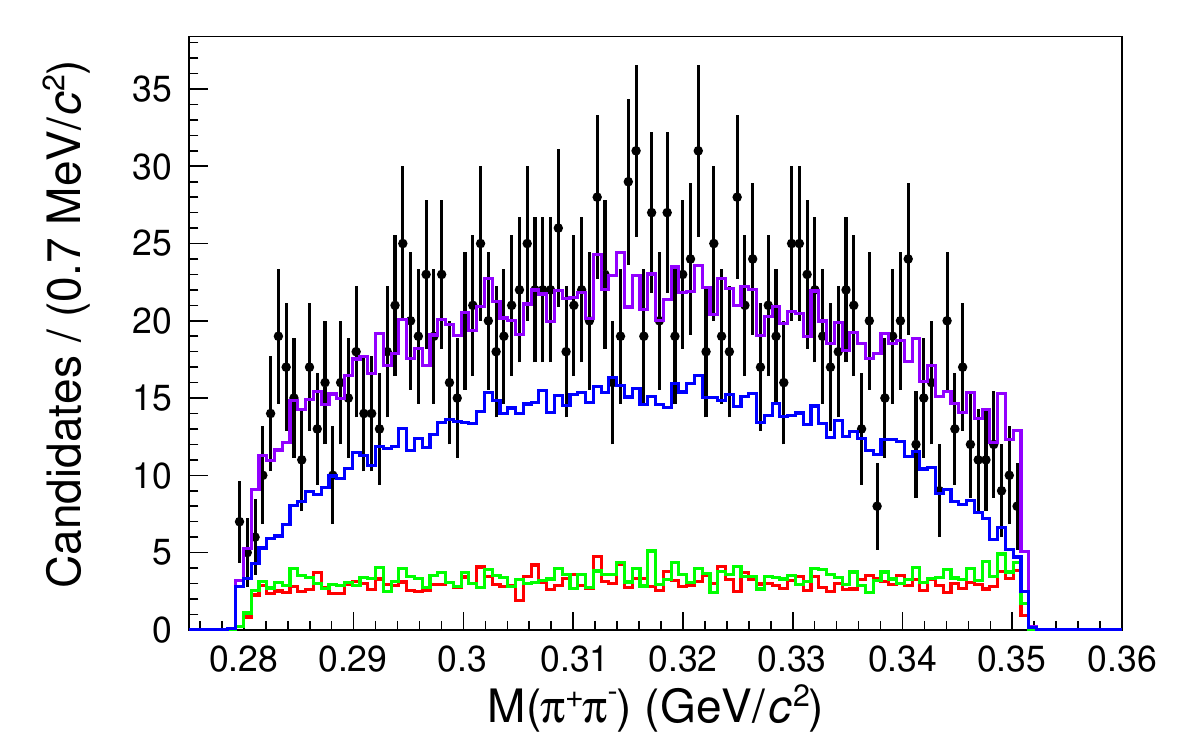} 
\includegraphics[width=0.48\linewidth]{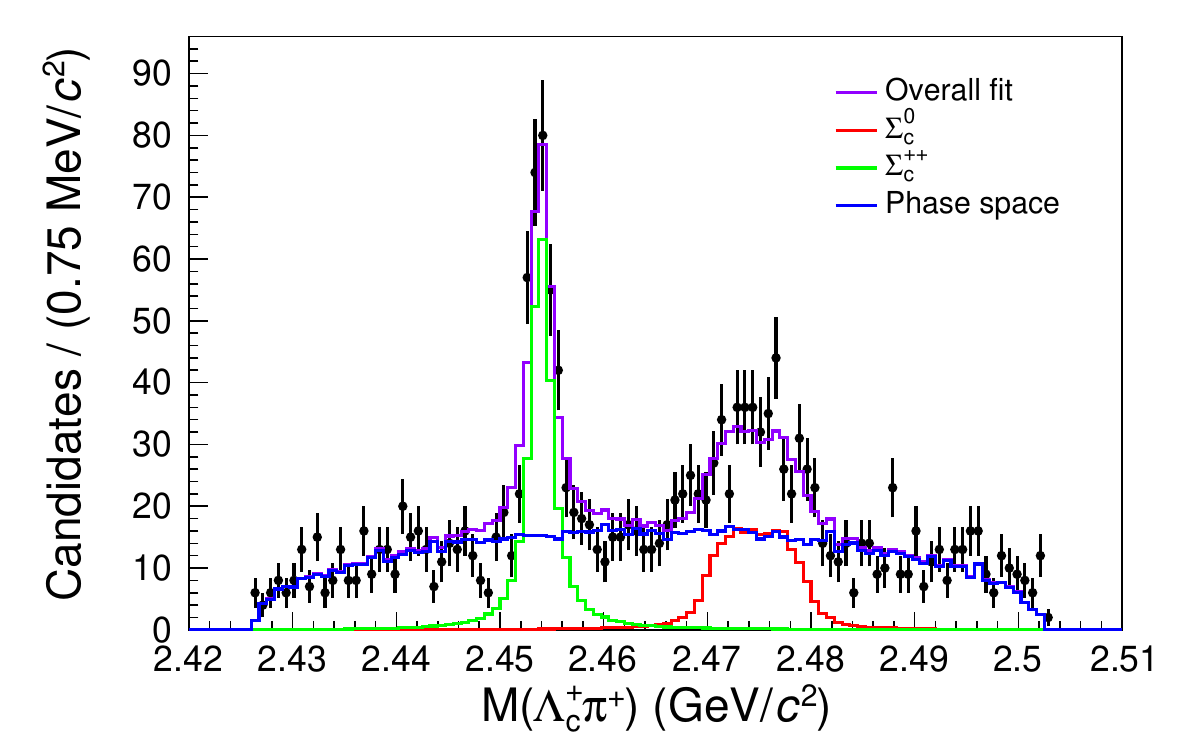}
\includegraphics[width=0.48\linewidth]{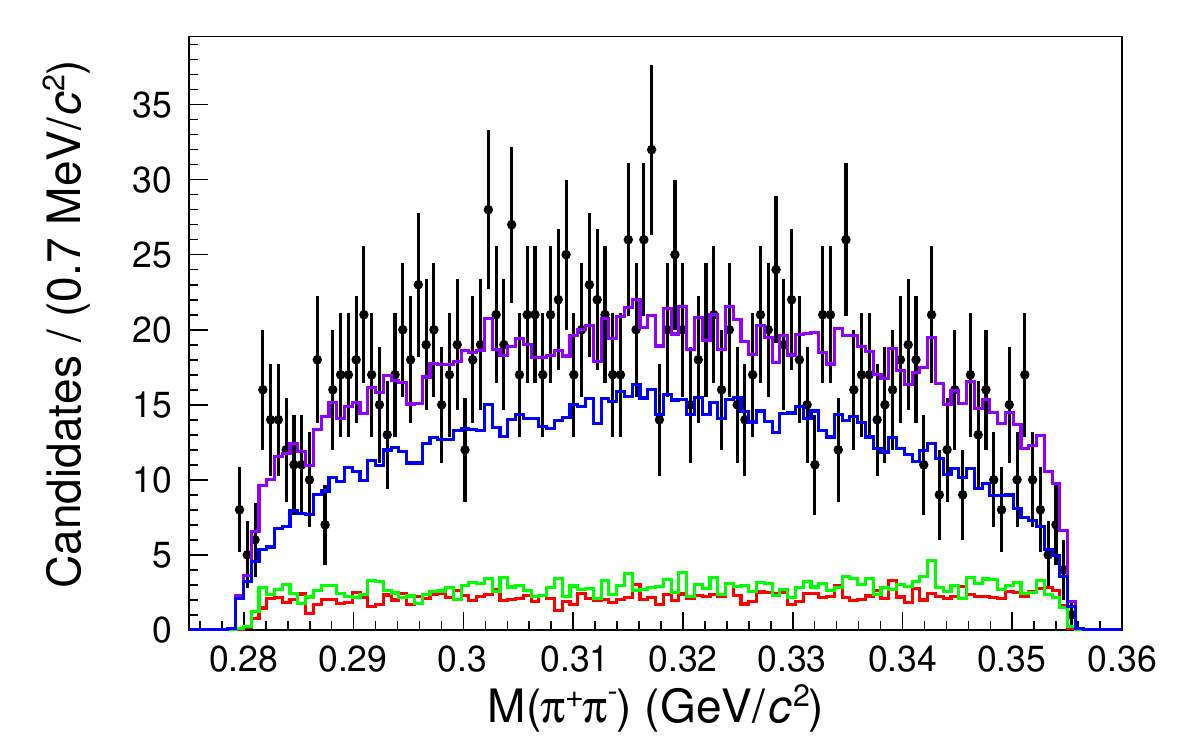} 
\includegraphics[width=0.48\linewidth]{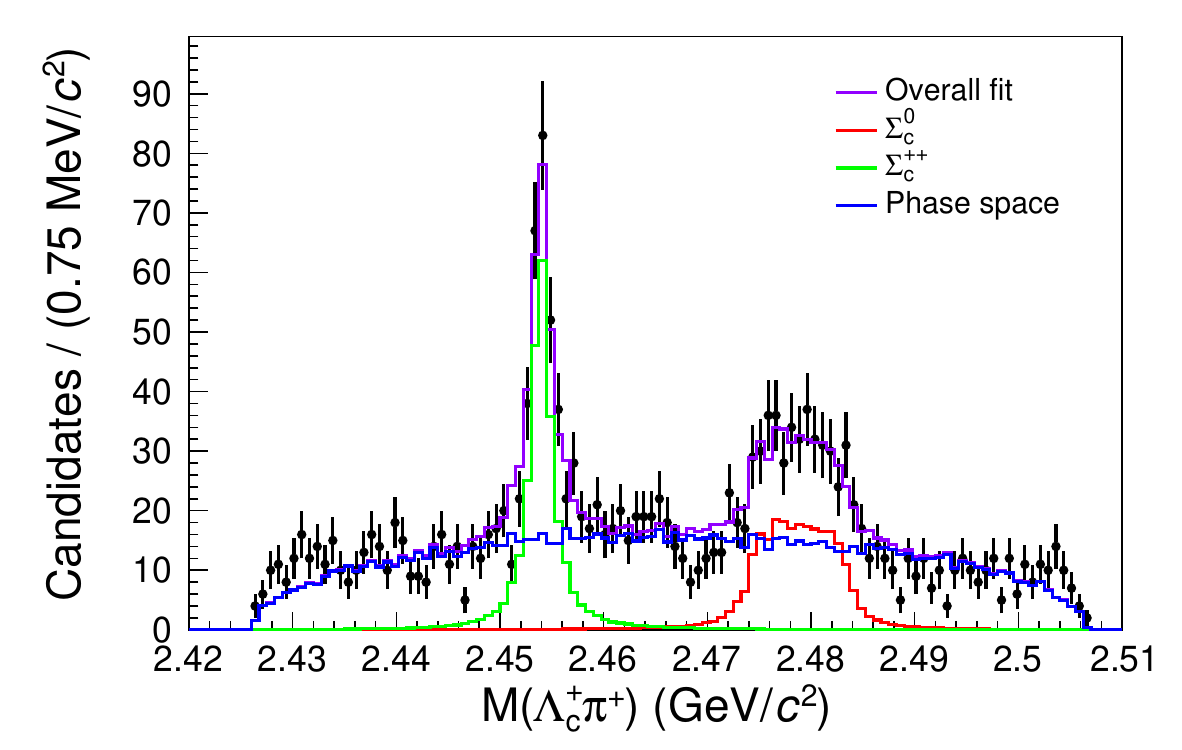}
\includegraphics[width=0.48\linewidth]{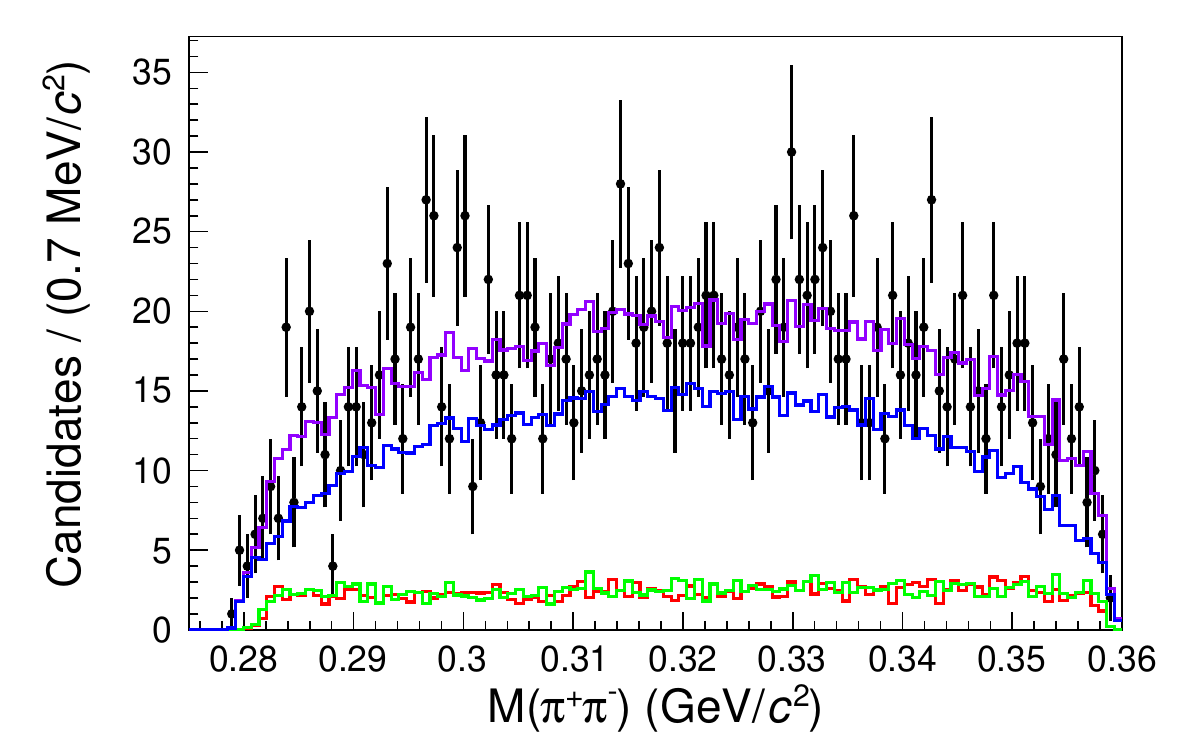} 
\caption{Projections of the Dalitz plot fits of the 3 sidebands on the right side of the signal region. The overall fitted distribution and the individual fitted components are shown alongside the experimental data.}
\label{fig:sideband_dalitz_right}
\end{figure*}

\begin{figure*}[htb]
\begin{subfigure}[b]{0.49\linewidth}
\includegraphics[width=\linewidth]{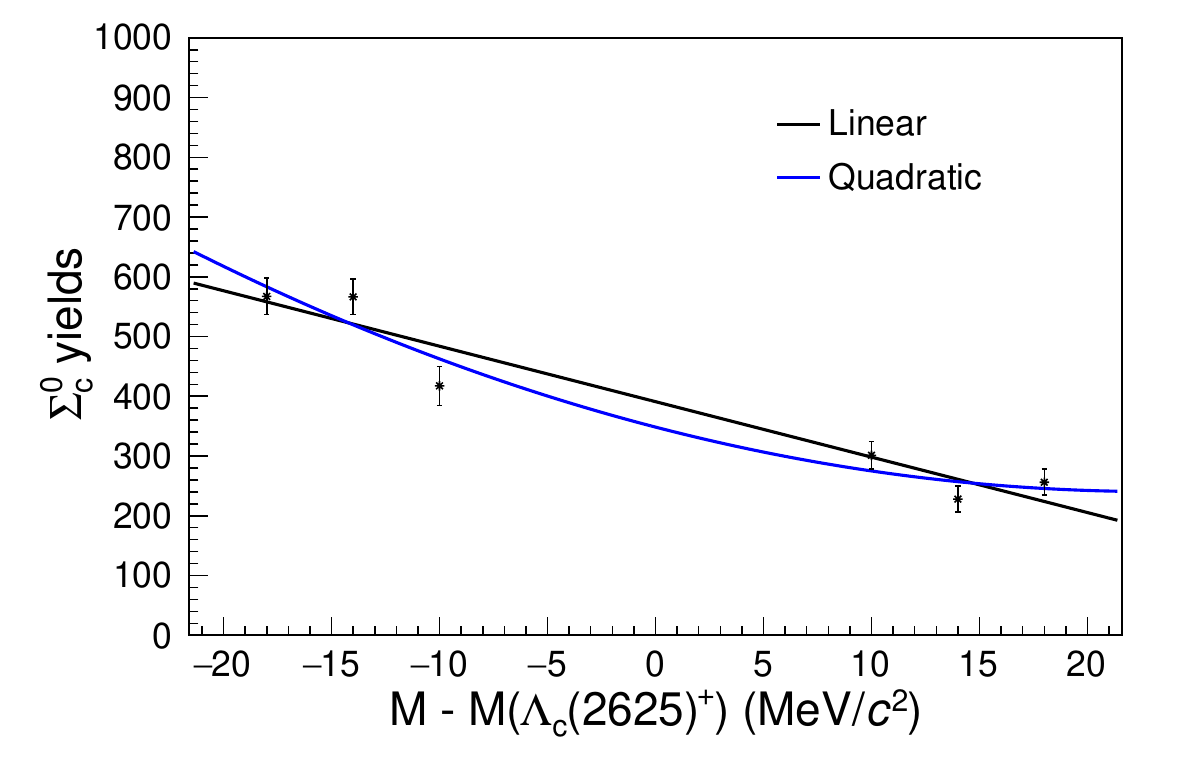}
\end{subfigure}
\begin{subfigure}[b]{0.49\linewidth}
\includegraphics[width=\linewidth]{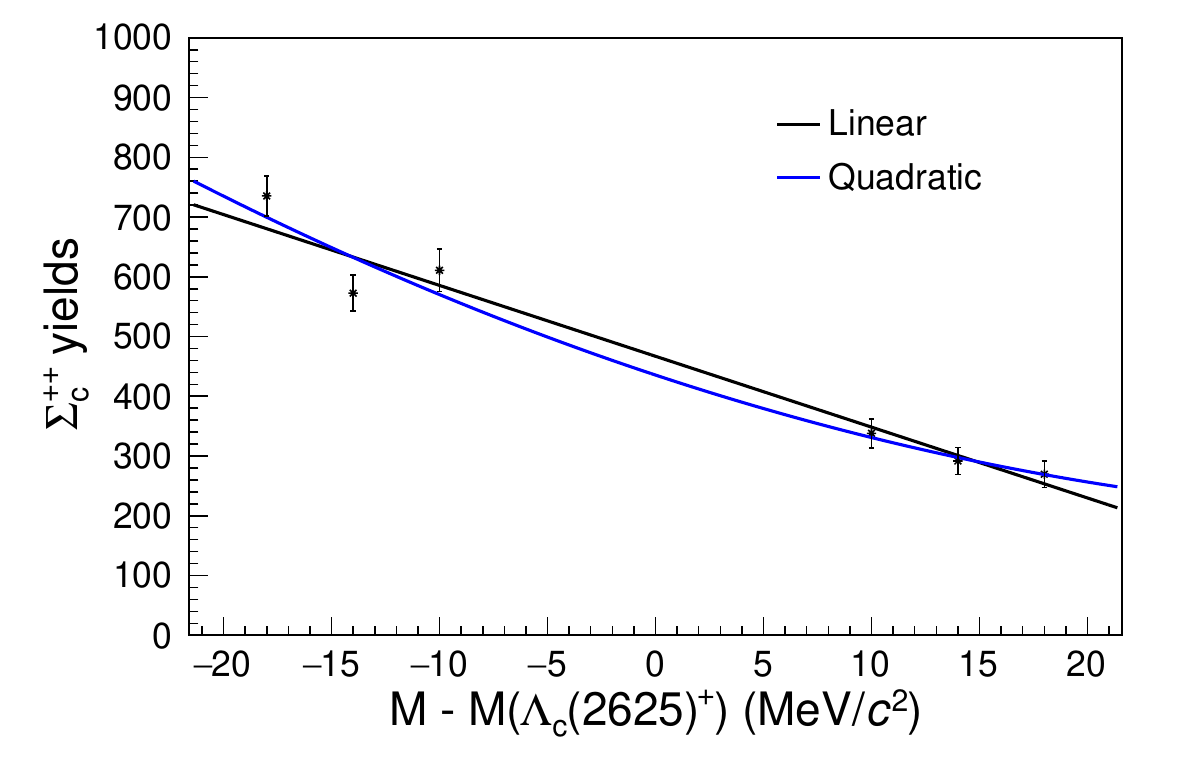}
\end{subfigure}
\caption{$\Sigma_c^0$ and $\Sigma_c^{++}$ yields from the 
sideband Dalitz plot fits, overlaid with linear and quadratic extrapolations.}
\label{fig:simgac_yields}
\end{figure*}

To account for $\Sigma_c(2455)^{++/0}$ candidates that 
are not decay products of the $\Lambda_c(2625)^+$,
the  $\Sigma_c(2455)^{++/0}$ yields from the
$M(\Lambda_c^+\pi^+\pi^-)$ sidebands are subtracted from 
the $\Sigma_c(2455)^{++/0}$ yields found from the amplitude fit.
The sidebands are six $4\,\mevcsq$ regions near the 
$\Lambda_c(2625)^+$ signal region, 
as shown in Fig.~\ref{fig:region_def}.
Each sideband region is fitted as 
an incoherent sum of the contributions from 
the $\Sigma_c(2455)^0$, the $\Sigma_c(2455)^{++}$, 
and the three-body phase space decay.
Figures~\ref{fig:sideband_dalitz_left} and 
\ref{fig:sideband_dalitz_right}
show the projections of the fit results for each
sideband region with each component labeled on the 
plot. 
The $\Sigma_c(2455)^{++/0}$ yields in the signal
region are determined by extrapolating the yields 
from the sidebands according to a linear fit, as shown
in Fig.~\ref{fig:simgac_yields} and tabulated in 
Table~\ref{tab:subtractedYields}.
The background yields to be subtracted 
are $N_{\mathrm{bkg}}(\Sigma_c^{0}) = 391 \pm 11$ and 
$N_{\mathrm{bkg}}(\Sigma_c^{++}) = 467 \pm 12$.
The branching ratio of 
$\Lambda_c(2625)^+ \to \Sigma_c^{0} \pi^{+}$
relative to the reference mode 
$\Lambda_c(2625)^+ \to \Lambda_c^+ \pi^{+} \pi^{-}$
is calculated using
\begin{equation} \label{eq:br_calc}
\begin{split}
\frac{\mathcal{B}(\Lambda_c(2625)^+ \to \Sigma_c^{0} \pi^{+})}
{\mathcal{B}(\Lambda_c(2625)^+ \to \Lambda_c^+ \pi^{+} \pi^{-})} 
= \frac{N_{\mathrm{sig}}(\Sigma_c^{0}) - 
N_{\mathrm{bkg}}(\Sigma_c^{0})}{N_{\mathrm{sig}}(\Lambda_c(2625)^+)}
\end{split}
\end{equation}
and similarly for the $\Sigma_c^{++}\pi^-$ mode.
We note that the efficiency over the area of the Dalitz plot is found to be uniform to within the statistical precision of the MC simulation.

We measure the branching ratios to be
\begin{equation}
\begin{aligned}
\frac{\mathcal{B}(\Lambda_c(2625)^+ \to \Sigma_c^{0} \pi^{+})}
{\mathcal{B}(\Lambda_c(2625)^+ \to \Lambda_c^+ \pi^{+} \pi^{-})} 
= (5.19 \pm 0.23)\% \\
\frac{\mathcal{B}(\Lambda_c(2625)^+ \to \Sigma_c^{++} \pi^{-})}
{\mathcal{B}(\Lambda_c(2625)^+ \to \Lambda_c^+ \pi^{+} \pi^{-})} 
= (5.13 \pm 0.26)\%
\end{aligned}
\end{equation}
where the errors are statistical only.

\section{Systematic uncertainties}

The primary source of systematic uncertainty on the
$\Lambda_c(2625)^+$ width is the inconsistency of the mass-resolution
function between the MC sample and the experimental data.
We use the $D^{*+}\to D^0\pi^+$ decay
as a control sample to determine the 
under- or over-estimation of the mass resolution in the MC sample
relative to the experimental data. 
The mass resolution in the experimental data is found to 
be 86\% of that in the MC sample with track smearing,
114\% without track smearing.
Both mass-resolution functions are used to determine the 
upper limit on the $\Lambda_c(2625)^+$ width in the 
experimental data. 
When applied to the experimental data, 
the mass resolution without track smearing scaled by 114\%
results in a more conservative upper limit on the 
$\Lambda_c(2625)^+$ width, thus reported as the 
final result.

The systematic uncertainty on the $\Lambda_c(2625)^+$ mass
is not greatly affected by the uncertainty on the 
$M(\Lambda_c^+\pi^+\pi^-)$ mass resolution,
but is limited by the precision with which the Belle detector
can measure the mass in this range.
Studies with $D^{*+} \to D^0\pi^+$ decays show that the 
measured $D^{*+}$ mass difference with respect to the
world-average value is $0.004 \,\mevcsq$. 
Any imperfection in the soft pion momentum calibration
changes the measured mass of the $\Lambda_c(2625)^+$ 
more than that of the $D^{*+}$.
We determine the scale factor required to correct
the soft pion momentum such that the $D^{*+}$ mass
matches its PDG value, then apply the same scale 
factor to the daughter pions from $\Lambda_c(2625)^+$ candidates.
The $\Lambda_c(2625)^+$ mass changed by $0.042 \,\mevcsq$,
which we assign as the systematic uncertainty due to 
the mass scale.
The track smearing correction applied to 
tracks in the MC sample
has a negligible effect on the mass measurement. 
The systematic uncertainty due to the low momentum 
correction is $0.025 \,\mevcsq$, which is found
by comparing the measured $\Lambda_c(2625)^+$ mass
with and without the low momentum correction.
Summing the contributions from the mass scale and 
low momentum track correction in quadrature, 
the total systematic uncertainty
on the mass measurement is $0.049 \,\mevcsq$.

The systematic uncertainties on the branching ratios,
which are calculated from Eq.~(\ref{eq:br_calc}),
are derived from the uncertainty of the $\Lambda_c(2625)^+$
yield in the signal region, the $\Sigma_c^{++/0}$ yields
fitted in the signal Dalitz plot fit, and the $\Sigma_c^{++/0}$
subtracted yields extrapolated from the sideband fits.
The systematic uncertainty associated with each is
tabulated in Table~\ref{tab:yields_uncertainty}.
The $\Lambda_c(2625)^+$ signal yield is most affected
by the mass-resolution function. We fit the experimental data
with a mass-resolution function determined with and without
track smearing. The difference in the yields is taken as
the systematic uncertainty on the 
$\Lambda_c(2625)^+$ signal yield. 
The $\Sigma_c^{++/0}$ signal yields are determined 
from the Dalitz plot fit, with their masses and widths
fixed to PDG values. The masses, widths, and mass resolutions
are systematically varied within the PDG uncertainties, and
the maximum change in the fitted $\Sigma_c^{++/0}$ yield
is taken as the systematic uncertainty.
In order to determine the sideband subtracted
yield, the six sidebands are fitted individually 
to determine the $\Sigma_c^{++/0}$ yields, 
with the yield and uncertainty in each sideband region
shown in Fig.~\ref{fig:simgac_yields}. The extrapolated
yield at the nominal $\Lambda_c(2625)^+$ mass 
is a weighted average of
the yields of the six sidebands.
We take the difference 
between the linear and quadratic extrapolation
as shown in Table~\ref{tab:subtractedYields}
as the systematic uncertainty due to the extrapolation.
To account for the statistical fluctuation due 
the finite MC sample sizes in the Dalitz plot fits,
we compare the fitted results using two statistically
independent MC samples of the same size. The difference
is taken as the systematic uncertainty due to the 
MC sample size.
With systematic uncertainties on the yields in
Eq.~(\ref{eq:br_calc}) listed in Table~\ref{tab:yields_uncertainty},
the total systematic uncertainties on the 
branching ratios are calculated from
the propagation of error and are listed in 
Table~\ref{tab:br_uncertainty}.
The total systematic uncertainty on the 
branching ratios
is found to be 0.40\% for the 
$\Sigma_c^0 \pi^+$ channel, and 
0.32\% for the $\Sigma_c^{++}\pi^-$ channel.


\begin{table}[htb]
\caption{Subtracted yields for $\Sigma_c^{++/0}$}
\label{tab:subtractedYields}
\begin{tabular}{@{\hspace{0.5cm}}l@{\hspace{0.5cm}}  @{\hspace{0.5cm}}c@{\hspace{0.5cm}}
@{\hspace{0.5cm}}c@{\hspace{0.5cm}}}
\hline \hline
Method & $\Sigma_c^{0}$ yield  
    & $\Sigma_c^{++}$ yield \\
\hline
Linear      & $391 \pm 11$ & $467 \pm 12$ \\
Quadratic   & $348 \pm 26$ & $436 \pm 28$ \\
\hline
Difference  & 11.00\%   & 6.64\% \\
\hline \hline
\end{tabular}
\end{table}

\begin{table*}[htb]
\caption{The percentage systematic uncertainties of the signal yields used for the 
branching ratio calculation.}
\label{tab:yields_uncertainty}
\begin{tabular}{@{\hspace{0.5cm}}l@{\hspace{0.5cm}}
@{\hspace{0.5cm}}c@{\hspace{0.5cm}}
@{\hspace{0.5cm}}c@{\hspace{0.5cm}}
@{\hspace{0.5cm}}c@{\hspace{0.5cm}}
@{\hspace{0.5cm}}c@{\hspace{0.5cm}}
@{\hspace{0.5cm}}c@{\hspace{0.5cm}}}
\hline \hline
Source & $\Sigma_c^{0}$ signal  & $\Sigma_c^{++}$ signal
    & $\Sigma_c^{0}$ sideband  & $\Sigma_c^{++}$ sideband
    & $\Lambda_c(2625)^+$ signal \\
\hline
Resolution                & 1.97\%  & 1.42\%  & 2.74\%  & 1.08\% & 3.64\%\\
$\Sigma_c^{0/++}$ width   & 4.00\%  & 2.26\%  & 2.52\%  & 2.38\% & -\\
$\Sigma_c^{0/++}$ mass    & 1.25\%  & 1.11\%  & 0.08\%  & 0.08\% & -\\
Extrapolation             & - & - &            11.00\% & 6.64\%  & - \\
MC sample size            & 1.91\%      & 2.09\%        & 0.71\% & 0.22\% & - \\
\hline
Total                     & 5.01\%      & 3.57\%  & 11.63\% & 7.14\% & 3.64\% \\
\hline \hline
\end{tabular}
\end{table*}

\begin{table*}[htb]
\caption{Systematic uncertainties on the branching ratios.}
\label{tab:br_uncertainty}
\begin{tabular}{@{\hspace{0.5cm}}l@{\hspace{0.5cm}}
@{\hspace{0.5cm}}c@{\hspace{0.5cm}}
@{\hspace{0.5cm}}c@{\hspace{0.5cm}}
}
\hline \hline
Source & 
$\frac{\mathcal{B}(\Lambda_c(2625)^+ \to \Sigma_c^{0} \pi^{+})}
{\mathcal{B}(\Lambda_c(2625)^+ \to \Lambda_c^+ \pi^{+} \pi^{-})}$ &
$\frac{\mathcal{B}(\Lambda_c(2625)^+ \to \Sigma_c^{++} \pi^{-})}
{\mathcal{B}(\Lambda_c(2625)^+ \to \Lambda_c^+ \pi^{+} \pi^{-})}$ \\
\hline
$\Sigma_c^{0/++}$ resolution &  0.13\% & 0.10\% \\
$\Sigma_c^{0/++}$ width      &  0.26\% & 0.16\% \\
$\Sigma_c^{0/++}$ mass       &  0.08\% & 0.07\% \\
Extrapolation                &  0.14\% & 0.10\% \\
MC sample size               &  0.12\% & 0.14\% \\
$\Lambda_c^+(2625)$ resolution & 0.19\% & 0.19\% \\
\hline
Total                        & 0.40\%  & 0.32\% \\
\hline \hline
\end{tabular}
\end{table*}

\section{Discussion}

We report the most precise $\Lambda_c(2625)^+$ mass, 
width, and branching
ratio measurements to date. 
The measured mass is consistent with previous results.
The measured upper limit on the $\Lambda_c(2625)^+$ width
is $\Gamma(\Lambda_c(2625)^+) < 0.52\,\mevcsq$ 
at the 90\% confidence level.
Theoretical predictions for the $\Lambda_c(2625)^+$ width vary.
Arifi {\em et al.} predict the width to be 0.570 MeV/$\textrm{c}^2$ based on 
chiral and heavy quark symmetry~\cite{Arifi:2018prd}.
The width is revised to be between 0.09 and 0.26 MeV/$\textrm{c}^2$ 
in a subsequent publication with the inclusion of
relativistic corrections~\cite{Arifi:2021prd}. 
Kawakami {\em et al.} predict a width in the range of 
0.11 - 0.73 MeV/$\textrm{c}^2$ based on chiral symmetry~\cite{Kawakami:2018prd}.
Guo {\em et al.} predict a much smaller width of
$1.13\times10^{-2}$ MeV/$\textrm{c}^2$,
based on the $^3P_0$ model~\cite{Guo:2019prd}.

%
The branching ratios of 
$\Lambda_c(2625)^+ \to \Sigma_c^0 \pi^+$ and 
$\Lambda_c(2625)^+ \to \Sigma_c^{++} \pi^-$
relative to the reference mode 
$\Lambda_c(2625)^+ \to \Lambda_c^{++} \pi^+ \pi^-$
are extracted from a full Dalitz plot fit.
Backgrounds from non-$\Lambda_c(2625)^+$ decays are
subtracted from the $\Sigma_c^{++/0}$ yields.
Our measurements align with the prediction 
by Arifi {\em et al.}, who assume $\Lambda_c(2625)^+$
is a $\lambda$ mode excitation~\cite{Arifi:2018prd}. 
Kawakami {\em et al.} predicted
a wide range~\cite{Kawakami:2018prd} and 
Guo {\em et al.} predicted the ratio 
$\Gamma(\Sigma_c^{++} \pi^{-})/\Gamma_{\mathrm{total}}$
to be 
29.9\%~\cite{Guo:2019prd}, 
which is already in contradiction with the previous measurement.
Our measurements of the properties of the $\Lambda_c(2625)^+$ charmed 
baryon will be useful to further constrain the parameter
space of the quark models and can be applied to other heavy quark systems.

\section{Conclusions}

We measure the mass of the $\Lambda_c(2625)^+$ to be 
$2627.978 \pm 0.006 \pm 0.049\,\mevcsq$, where 
the uncertainty on the $\Lambda_c^+$ mass is not included
since it is constrained to the PDG value during reconstruction.
This is equivalent to
\begin{equation}
M(\Lambda_c(2625)^+) - M(\Lambda_c^+) =
341.518 \pm 0.006 \pm 0.049 \,\mevcsq
\end{equation}
The mass measurement is consistent with
the previous CDF measurement but with approximately
half the uncertainty
\cite{CDF:2011}.

An upper limit on the $\Lambda_c(2625)^+$ width is determined to be
\begin{equation}
\Gamma(\Lambda_c(2625)^+) < 0.52 \,\mevcsq\
\end{equation}
at 90\% confidence level 
which is around a factor of two more stringent than the previous
limit.

Based on a full Dalitz plot fit and with sideband subtraction 
of the $\Sigma_c^{++/0}$ yields,
the branching ratios
relative to the mode $\Lambda_c(2625)^+ \to \Lambda_c^+ \pi^+ \pi^-$
are obtained:

\begin{equation}
\begin{aligned}
\frac{\mathcal{B}(\Lambda_c(2625)^+ \to \Sigma_c^{0} \pi^{+})}
{\mathcal{B}(\Lambda_c(2625)^+ \to \Lambda_c^+ \pi^{+} \pi^{-})} = 
(5.19 \pm 0.23 \pm 0.40)\% \\
\frac{\mathcal{B}(\Lambda_c(2625)^+ \to \Sigma_c^{++} \pi^{-})}
{\mathcal{B}(\Lambda_c(2625)^+ \to \Lambda_c^+ \pi^{+} \pi^{-})} 
= (5.13 \pm 0.26 \pm 0.32) \%
\end{aligned}
\end{equation}
This is the first measurement made of 
these branching ratios as previously only limits have been presented.
These measurements can be used as inputs to theoretical models
to generate predictions for other heavy quark baryons.

\section{Acknowledgements}
This work, based on data collected using the Belle detector, which was
operated until June 2010, was supported by 
the Ministry of Education, Culture, Sports, Science, and
Technology (MEXT) of Japan, the Japan Society for the 
Promotion of Science (JSPS), and the Tau-Lepton Physics 
Research Center of Nagoya University; 
the Australian Research Council including grants
DP180102629, 
DP170102389, 
DP170102204, 
DE220100462, 
DP150103061, 
FT130100303; 
Austrian Federal Ministry of Education, Science and Research (FWF) and
FWF Austrian Science Fund No.~P~31361-N36;
the National Natural Science Foundation of China under Contracts
No.~11675166,  
No.~11705209;  
No.~11975076;  
No.~12135005;  
No.~12175041;  
No.~12161141008; 
Key Research Program of Frontier Sciences, Chinese Academy of Sciences (CAS), Grant No.~QYZDJ-SSW-SLH011; 
Project ZR2022JQ02 supported by Shandong Provincial Natural Science Foundation;
the Ministry of Education, Youth and Sports of the Czech
Republic under Contract No.~LTT17020;
the Czech Science Foundation Grant No. 22-18469S;
Horizon 2020 ERC Advanced Grant No.~884719 and ERC Starting Grant No.~947006 ``InterLeptons'' (European Union);
the Carl Zeiss Foundation, the Deutsche Forschungsgemeinschaft, the
Excellence Cluster Universe, and the VolkswagenStiftung;
the Department of Atomic Energy (Project Identification No. RTI 4002) and the Department of Science and Technology of India; 
the Istituto Nazionale di Fisica Nucleare of Italy; 
National Research Foundation (NRF) of Korea Grant
Nos.~2016R1\-D1A1B\-02012900, 2018R1\-A2B\-3003643,
2018R1\-A6A1A\-06024970, RS\-2022\-00197659,
2019R1\-I1A3A\-01058933, 2021R1\-A6A1A\-03043957,
2021R1\-F1A\-1060423, 2021R1\-F1A\-1064008, 2022R1\-A2C\-1003993;
Radiation Science Research Institute, Foreign Large-size Research Facility Application Supporting project, the Global Science Experimental Data Hub Center of the Korea Institute of Science and Technology Information and KREONET/GLORIAD;
the Polish Ministry of Science and Higher Education and 
the National Science Center;
the Ministry of Science and Higher Education of the Russian Federation, Agreement 14.W03.31.0026, 
and the HSE University Basic Research Program, Moscow; 
University of Tabuk research grants
S-1440-0321, S-0256-1438, and S-0280-1439 (Saudi Arabia);
the Slovenian Research Agency Grant Nos. J1-9124 and P1-0135;
Ikerbasque, Basque Foundation for Science, Spain;
the Swiss National Science Foundation; 
the Ministry of Education and the Ministry of Science and Technology of Taiwan;
and the United States Department of Energy and the National Science Foundation.
These acknowledgements are not to be interpreted as an endorsement of any
statement made by any of our institutes, funding agencies, governments, or
their representatives.
We thank the KEKB group for the excellent operation of the
accelerator; the KEK cryogenics group for the efficient
operation of the solenoid; and the KEK computer group and the Pacific Northwest National
Laboratory (PNNL) Environmental Molecular Sciences Laboratory (EMSL)
computing group for strong computing support; and the National
Institute of Informatics, and Science Information NETwork 6 (SINET6) for
valuable network support.

\bibliography{paper}

\begin{thebibliography}{22}
\bibitem [{cha()}]{charge_conjugates}%
  \BibitemOpen
  \href@noop {} {}\bibinfo {note} {{Reconstruction of charge conjugates is
  implied unless otherwise stated}}\BibitemShut {NoStop}%
\end{thebibliography}

\end{document}